
The MS1224+20 cluster of galaxies is a high luminosity X-ray source at
z=0.325.  To compare the lensing mass to the virial mass we have
completed a uniform, high precision redshift survey over a field of
$7\arcm\times9\arcm$, obtaining 75 redshifts.  The velocity dispersion
of 30 cluster galaxies is $775$ \kms\ and the projected harmonic
radius is 0.32 \hmpc.  The virial mass is $2.1\times 10^{14} h^{-1}
\msun$.  Correcting for faint cluster galaxies without redshifts and
allowing for a modest evolution in the galaxy luminosity function
gives a current epoch mass-to-light ratio $M/L_V(0)=255h \msun/\lsun$.
The same field contains a $z=0.225$ cluster with $\sigma_v\simeq 500$
\kms, and a group at $z=0.412$ with $\sigma_v\simeq 400$ \kms.  The
clusters' gravitational field induces image ellipticities that are
calculated from the light-traces-mass density distribution and
compared to the observed average tangential distortions from Fahlman
\et (1994).  Between 1.5 and 3 virial radii, the observed lensing
distortions are $4\pm2$ times stronger than the light-traces-mass
model and the virial M/L predict.
\\

\newif\ifapj
\ifapj
	\magnification=\magstep1
\else
	\magnification=\magstephalf
\fi
\input epsf
\epsfverbosetrue
\font\caps=cmcsc10
\font\bb=cmbx12
\tolerance=10000 \pretolerance=5000
\looseness=-3
\widowpenalty 2000
\displaywidowpenalty 2000
\overfullrule 0pt

\ifapj\baselineskip 24pt
 \else \baselineskip=15truept
\fi
\parindent=2.5em
\def\ub{\underbar}
\def\hi{\par\noindent \hangindent=2.5em}

\newcount\eqnumber
\eqnumber=1
\def\eqnum{\the\eqnumber\global\advance\eqnumber by 1}
\def\ls{\vskip 12.045pt}
\def\ni{\noindent}
\def\et{{\it et\thinspace al.}\ }
\def\eg{{\it e.~g.}\ }
\def\kms{km\thinspace s$^{-1}$ }
\def\deg{\ifmmode^\circ\else$^\circ$\fi}    

\def\msun{M_\odot}
\def\lsun{L_\odot}
\def\arcs{\ifmmode {'' }\else $'' $\fi}     
\def\arcm{\ifmmode {' }\else $' $\fi}     
\def\buildrel#1\over#2{\mathrel{\mathop{\null#2}\limits^{#1}}}
\def\mper{\ifmmode \buildrel m\over . \else $\buildrel m\over .$\fi}
\def\hper{\ifmmode \rlap.^{h}\else $\rlap{.}^h$\fi}
\def\sper{\ifmmode \rlap.^{s}\else $\rlap{.}^s$\fi}
\def\arcsper{\ifmmode \rlap.{' }\else $\rlap{.}' $\fi}
\def\arcmper{\ifmmode \rlap.{'' }\else $\rlap{.}'' $\fi}

\def\aj{{AJ}, }
\def\apj{{ApJ}, }
\def\apjs{{ApJSup}, }
\def\apjl{{ApJLett}, }

\def\mn{{MNRAS}, }

\def\spose#1{\hbox to 0pt{#1\hss}}
\def\lta{\mathrel{\spose{\lower 3pt\hbox{$\mathchar"218$}}
     \raise 2.0pt\hbox{$\mathchar"13C$}}}
\def\gta{\mathrel{\spose{\lower 3pt\hbox{$\mathchar"218$}}
     \raise 2.0pt\hbox{$\mathchar"13E$}}}

\def\hmpc{$h^{-1}$\thinspace Mpc}

\ifapj\vglue 29.10pt \fi

\centerline{\bb LENSING FROM THE LIGHT-TRACES-MASS MAP OF MS1224+20}
\vskip 1.0truecm
\noindent
\centerline{\bf R. G. Carlberg$^{1,2,3}$, H. K. C. Yee$^{1,2}$,
	and E. Ellingson$^{1,4}$}
\vskip 1truecm
\centerline{$^2$Department of Astronomy, University of Toronto,}
\centerline{60 St.~George St., Toronto, Ontario, M5S 1A1, Canada}
\medskip
\centerline{$^3$Department of Astronomy, FM-20, University of Washington,}
\centerline{Seattle, WA 98195}
\medskip
\centerline{$^4$ Center for Astrophysics and Space Astronomy, CB 389, }
\centerline{University of Colorado, Boulder 80309}
\footnote{}{\noindent $^1$ Guest observer,
Canada-France-Hawaii Telescope, operated jointly by CNRS of
France, NRC of Canada, and the University of Hawaii}

\bigskip\bigskip
\ifapj
\centerline{received: $\underline{\hbox to 6truecm{\hphantom\hfill}}$}
\vskip 0.5truein
\centerline{accepted: $\underline{\hbox to 6truecm{\hphantom\null}}$}
\vfill\eject
\fi

\centerline{ABSTRACT}

The MS1224+20 cluster of galaxies is a high luminosity X-ray source at
z=0.325.  To compare the lensing mass to the virial mass we have
completed a uniform, high precision redshift survey over a field of
$7\arcm\times9\arcm$, obtaining 75 redshifts.  The velocity dispersion
of 30 cluster galaxies is $775$ \kms\ and the projected harmonic
radius is 0.32 \hmpc.  The virial mass is $2.1\times 10^{14} h^{-1}
\msun$.  Correcting for faint cluster galaxies without redshifts and
allowing for a modest evolution in the galaxy luminosity function
gives a current epoch mass-to-light ratio $M/L_V(0)=255h \msun/\lsun$.
The same field contains a $z=0.225$ cluster with $\sigma_v\simeq 500$
\kms, and a group at $z=0.412$ with $\sigma_v\simeq 400$ \kms.  The
clusters' gravitational field induces image ellipticities that are
calculated from the light-traces-mass density distribution and
compared to the observed average tangential distortions from Fahlman
\et (1994).  Between 1.5 and 3 virial radii, the observed lensing
distortions are $4\pm2$ times stronger than the light-traces-mass
model and the virial M/L predict.

\vfill\eject

\ni\ub{1. INTRODUCTION}

\ls
The possibility that the total mass of galaxy clusters may be far
larger than the virial masses calculated from their galaxy
distributions has long been recognized, but is an observationally
difficult problem. The virial mass statistic is not sensitive to mass
beyond the mean harmonic radius of the (effectively massless) galaxies
which trace the cluster potential.  Cluster virial mass-to-light
ratios (\eg Kent and Gunn 1982) have relatively stable values implying
that $\Omega\simeq0.2$, in close agreement with those inferred from
the cosmic virial theorem (Bean \et 1983, Davis \& Peebles
1983). However, large scale flow fields measure $\sigma_8\Omega^{0.6}$
and generally are taken to imply that $\Omega\simeq 1$, with large
confidence intervals and uncertainty as to the (sample dependent)
value of $\sigma_8$, the 8\hmpc\ sphere density contrast (\eg
Lynden-Bell \et 1989, Bertschinger \et 1990, Kaiser \et 1991, Strauss
\et 1992, Nusser \& Dekel 1993).  The discrepancy between these
measurements of $\Omega$ is somewhat surprising since clusters of
galaxies are the largest collapsed structures, and the length scales
from which they originate are only a factor of three smaller than the
scales on which the flows are measured. Aside from the observational
issue of establishing $\Omega$,
the determination of cluster masses
is also a significant test
inflationary cosmology, which in its simplest form predicts $\Omega=1$
(Guth 1981, Bardeen \et 1983).  Some physical motivation for a
galaxy-mass segregation of approximately the correct magnitude to
allow cluster galaxies to indicate $\Omega\simeq 0.2$ even though the
true global value is unity is found within (some, not all) numerical
simulations following the infall of galaxies into growing clusters
(\eg West \& Richstone 1988, Carlberg
\& Dubinski 1991, Carlberg 1994).
Observational evidence that virial masses underestimate
total cluster masses or that light does not trace mass
would be an important step in resolving
these issues.

To measure galaxy cluster masses at radii well beyond the virial
radius of course requires mass tracers at radii where the cluster
density is rapidly declining (faster than $r^{-2}$ according to
simulation data). Galaxies themselves remain viable tracers, although
the problem of unrelated field galaxies in the the velocity space of
the cluster becomes an important issue.
X-ray flux diminishes as the square of the density
making the measurements difficult.
The relatively new technique of weak gravitational lensing (\eg Tyson \et
1990, Miralda-Escude 1991) has great promise as an indicator of
the total mass and does not depend on the orbital complications of
that mass, but is subject to some uncertainties in the practical
details of the method, such as the redshifts of the lensed galaxies.

In this paper we use a redshift survey of the MS1224+20 cluster field to
construct a light-traces-mass surface density map normalized using the
mass-to-light ratio ($M/L$) from a virial analysis of the cluster.
{}From the mass map the weak gravitational lensing distortions are
calculated, and compared to the distortions observed by Fahlman \et
(1994). Our basic goal is to test whether the mass detected by lensing
is consistent with the virial mass.  The next section discusses our
observations, which we use in \S3 to estimate the virial mass; and in
\S4 the gravitational distortions, which we compare to the observed
values.

\bigskip\goodbreak
\ni\ub{2. OBSERVATIONS}

\ls
The cluster MS1224+20 (Gioia \et 1990, Henry \et 1992, Gioia \&
Luppino 1994) has a very high X-ray luminosity ($L_x\simeq 4.6\times
10^{44}$ erg~s$^{-1}$) at z=0.325 and is one of a uniform sample of
moderate redshift clusters for which we doing a high precision
redshift survey. The cluster was observed at CFHT during the nights UT
1993 Jan 24 and 25 using the Multi-Object Spectrograph (MOS) as part
of the CNOC (Canadian Network for Observational Cosmology) consortium
redshift survey of galaxy clusters (see Carlberg \et 1994).  The
cluster field (useful size $430\arcs\times540\arcs$, centered on the
cD galaxy) was imaged in Gunn $r$ and $g$ filters for 900 seconds
each.  Using a variable point-spread function version of the faint
galaxy photometry software system PPP (Yee 1991), two-color photometry
and star-galaxy separation of all objects in the field was performed
in real time. The resulting catalog was then used as input to a
program which optimally designed spectroscopic aperture masks for the
MOS.  For a detailed description of our observational techniques see
Yee \et (1994).
In order to obtain as many spectra per aperture mask as possible,
a filter was used which limited the spectra to
4650--6100\AA. This range
passes the [OII] emission feature at 3727\AA, the 4000\AA\ break region,
and G-band (4304\AA) for objects in the
$0.25\lta z\lta 0.42$ range. This is optimal for identifying cluster
members at z=0.32, but introduces a strong redshift selection function
for field galaxies.
In particular,
emission line redshifts with $0.20 < z < 0.25$ are
unlikely to be obtained,
(as neither [OII] 3727\AA\ nor [OIII] 4959/5007\AA\ are observable),
although we have identified a number of absorption line galaxies with z=0.22
in the field. Likewise, absorption line galaxies with $z > 0.40$ are
more difficult to identify.
The nominal completeness limit for spectroscopy was set at $r=21.5$ mag
with a spectral resolution of 15 \AA.
Two masks, each  with over 100 slits, were observed.
The integration time for
mask A was $2\times 2400$ seconds and mask B, $2\times 3600$ seconds,
yielding a total of 75 redshifts in
the field.
There are 30
redshifts in the cluster, and a total of 47 are associated with the 3
largest structures in the field (see Figures~1 and 2).
The velocity uncertainty of each redshift is estimated  by
multiple observations of several galaxies and by simulations
of galaxy spectra.

The observational procedures are designed to maximize the data
rate, maintain a velocity accuracy of better than $\simeq$150 \kms, and
preserve the geometric distribution of the galaxies in
the field (in particular the angular correlation function, and
the mean density of galaxies over the face of the cluster).
A completeness function $C(m,x,y)$, as a function of magnitude
and position, is estimated by
ratioing the total number of galaxies
to the number for which redshifts were obtained.
Due to the sparseness of data, the function is determined
separately in magnitude and position.
$C(m)$, derived using galaxies in the whole field,
 declines smoothly from 0.5 at $r=20$ mag, to 0.1 at $r=22$ mag, where
we truncate the sample. The entire $r=17$ to 22 mag
range is added together
to determine the geometrical selection effects, $C(x,y)$. The variance
of the selection function over the field is about 10\%.
All statistical quantities are geometrically weighted, which
produces only a small correction to the unweighted data.

\bigskip\goodbreak
\ni\ub{3. VIRIAL ANALYSIS}

\setbox11=\vbox{
\noindent \ifapj\else\narrower\baselineskip 12pt\fi
Figure 1:\hskip 5mm The redshifts plotted against the relative locations
(in physical distance) in the RA direction with East at the
top. There are 8 more redshifts at $z>0.42$.}
\ifapj\else\midinsert
\epsfxsize=\hsize
\centerline{\epsfbox[48 360 588 588]{zE1224.ps}}
\copy11 \endinsert \fi

\ls
The MS1224+20 field has a somewhat surprising amount of projected
structure (given that
X-ray selection reduces the contaminating effects),
in particular a foreground cluster at $z=0.22$ with
a velocity dispersion of 500 \kms, and a small cluster  with
a velocity dispersion of about 400 \kms at $z=0.41$.
The main cluster is taken to span the redshift
range from 0.318 to 0.331, with an average redshift
of 0.3257. The velocity histogram has some
internal structure (significant at the 95\% level). This level
of substructure is consistent with all other clusters in our survey,
and is to be expected since clusters are subject to
constant gravitational infall.

The line-of-sight velocity dispersion of the cluster is $\sigma_v\simeq775$
\kms, and
the harmonic radius of the galaxies,
$r_h$, is 112\arcs,
which is about
half of the field size in the smallest direction.  The density of the
cluster is falling rapidly beyond this radius,
with a projected slope
$d\log{\Sigma_L}/d\log{r}=-1.5$, so that the harmonic radius is not
expected to be affected by the finite field size. The physical length
subtended by the mean harmonic radius is 0.32\hmpc. The implied virial
mass, $3\pi \sigma_{v}^2 r_h/2G$ is $2.12\times 10^{14}\msun$.
The rest frame luminosity of the cluster to $r=22$ mag
is $L_V=8.2\times 10^{11}\lsun$,
using
$V-r=0.17$ mag (Sebok 1986), and a K correction of $K_V=-0.60$ mag
(Coleman, Wu, \& Weedman 1980).
The color and K-correction are based on an average of elliptical
galaxy and early-type spirals.
The completeness correction
based on $C(m,x,y)$ for galaxies not observed spectroscopically has
increased the luminosity 1.9 times over the observed galaxies.  The
rest frame mass-to-light ratio $M/L_V=250h$ in solar units to this
magnitude limit.  To estimate the current epoch $M/L$ we approximate
the luminosity evolution as $\Delta M \simeq z$.  The light missed
below our magnitude limit, $0.16L_\ast$, is calculated to be an
additional 34\%, assuming that the luminosity function can be
represented as a Schechter function with $\alpha=1.25$. The fully
corrected $M/L_V(0)\simeq 255h$, which is similar to the $275h$ found
for Coma ( Kent \& Gunn 1982).  A bootstrap analysis (for the entire
sample, not only the cluster galaxies) finds that the $1\sigma$ error
of $M/L_V(0)$ is 30\%, with less than 0.1\% probability that
$M/L_V(0)$ exceeds 500h.

The two other structures in the field, at redshifts 0.22 and 0.412
have, respectively, only 12 and 7 spectroscopic redshifts giving
velocity dispersions of 500 \kms\ and 400 \kms. Both
have M/L values consistent with that obtained for the 0.33 cluster
(but with substantial uncertainties).

\setbox12=\vbox{
\noindent \ifapj\else\narrower\baselineskip 12pt\fi
Figure 2:\hskip 5mm The ``$\kappa$ map''
(the sum of $\kappa=\Sigma/\Sigma_c(z_l,z_s)$, with
lensed sources assumed to be at $z_s=0.6$)
of the MS1224+20 field. The $\kappa$ map  increases the weight of
low redshift structures. The galaxy luminosities are converted
to masses using $M/L_V(0)=255h$ and smoothed with a 30\arcs\ Gaussian
filter. The contours are at 0.01 increments in $\kappa$, and the
``sticks'' are proportional to the local image distortion. The
largest distortion is $\gamma=0.05$.}

\ifapj\else\midinsert
\epsfysize=4.0truein
\centerline{\epsfbox[24 240 588 660]{map.ps}}
\copy12 \endinsert \fi

\ls\goodbreak
\ni\ub{4. THE LIGHT-TRACES-MASS MAP AND WEAK LENSING}

\ls
To test the assumption that the cluster light traces the mass in
the cluster, and in particular whether the cluster light is more
concentrated than the cluster mass we use the observed galaxy
distribution and the $M/L$ ratio to estimate the surface density
of the cluster. The 47 relevant redshifts will of course
only be good for a fairly noisy image. These same galaxies are used
in Fahlman \et (1994) for their comparison to their estimate
of lensing surface density.
Note that the resulting mass map does not depend on
the photometric K corrections. The surface density map, smoothed with
a 30\arcs\ Gaussian filter is displayed in Figure~2. The contours are
linear in mass density.

The predicted image distortions (in the weak field limit) are the
``sticks'' in Figure~2, The distortions are calculated beginning with
the 2D Poisson equation, $\nabla^2\phi=2\sum \kappa$, where the sum is
over the 3 structures, and $\kappa=\Sigma/\Sigma_c(z_l,z_s)$.
The inverse critical
surface
density, in angular co-ordinates, for a structure at
redshift $z$ is $(\Sigma^\theta_c)^{-1}= 4\pi Gc^{-2}
(1+z_l)r_l^{-1}(1-r_l/r_s)$i,
where the subscript $l$ and $s$ denote the lens and source redshift.
For $\Omega=1$ the co-moving co-ordinate distances are
$r(z)=2cH_0^{-1}(1-(1+z)^{-1/2})$.
The total image
distortion is $\sqrt{\gamma_1^2+\gamma_2^2}$, where
$\gamma_1=(\phi_{xx}-\phi_{yy})/2$, $\gamma_2=\phi_{xy}$ (\eg
Schneider \et 1992). The distortion is oriented at an angle
$\tan{\varphi}=(\gamma_1-\gamma)/\gamma_2$. The tangential ellipticity
is defined as $e_T=\gamma(\cos{\varphi}\sin{2\theta}
-\sin{\varphi}\cos{2\theta})$, where $\theta$ is the polar co-ordinate
angle. For a circularly symmetric mass distribution the $\gamma$ values
are $2\overline{\Sigma}(<r) -\Sigma(r)$ in critical units, where
$\overline{\Sigma}$ is the mean interior mass density, $M/\pi r^2$.
Asymmetrically placed masses outside $r$ introduce additional terms.

The sources are taken to be at
$z_s=0.6$, in common with Fahlman \et (1994). This is an important
assumption (for which there is strong supporting evidence)
because for a given measured shear the lensing mass decreases for larger
redshift sources. The redshifts of cluster and background imply
a critical surface density of $7.5\times10^{15}\msun$~Mpc$^{-2}$.
At a projected distance of 1\hmpc\ from the cluster, i.e. about
3 virial radii, the
virial mass averaged over this radius
implies a density of 0.089 in critical units, which is the
approximate amplitude of the expected distortions. At the virial
radius the distortions should be about 10\%, but the jumble of structure
visible in Figure~2 will lead to a substantial noise in
the $\langle e_T \rangle$ measurement.

\setbox13=\vbox{
\noindent \ifapj\else\narrower\baselineskip 12pt\fi
Figure 3:\hskip 5mm
The predicted (line) and observed (squares) mean tangential
ellipticity plotted against the distance from the cD galaxy. The
error flags on the observed points are the errors of the mean, the
dashed lines show
the  variance of the predicted mean. The background sources
are taken to be at $z_s=0.6$.
The observed and predicted
tangential ellipticities are
in general agreement within 150\arcs. There is
a  2 standard deviation
significant excess of lensing distortions over the virial $M/L$ normalized
distortions predicted between 150 and 300 \arcs.
}
\ifapj\else\midinsert
\epsfysize=3.0truein
\centerline{\epsfbox[24 200 588 688]{dis.ps}}
\copy13 \endinsert \fi

A straightforward comparison between the mean background galaxy
distortions observed by Fahlman \et (1994) and those predicted from
the light distribution is to azimuthally average about some chosen
center. We use the cD galaxy as the center of the light, which is also
the center of the X-ray distribution (Gioia \et 1990).  Figure~3
displays the averaged tangential ellipticities predicted from the
light distribution (the solid line) and the observed values (the
squares). The predicted values have the upper and lower width of the
distribution shown as dashed lines, the observed values have error
flags that give the error of the mean. The main result is that the
observed lensing is considerably stronger than predicted by the virial
$M/L$ ratio.  Within the central 150\arcs (the virial radius is
112\arcs) the difference between the observed and predicted values is
not significant, but this is totally dependent on the choice of the
center of symmetry. Using the center of lensing mass gives much larger
observed $\langle e_T \rangle$.  Between 150 and 300\arcs\ the
observed average tangential image distortion ($\langle
e_T\rangle=0.061\pm0.025$) is about $4\pm2$ times above the average
predicted value, 0.010. Therefore we reject at the 2 standard
deviation level the hypothesis that the total mass enclosed at 2
virial radii is the virial mass.

\bigskip\goodbreak
\ni\ub{5. DISCUSSION AND CONCLUSIONS}

\ls
Combining a dynamical analysis of visible galaxies and the weak
distortion map of background galaxies allows a powerful test of the
hypothesis that cluster mass distributions are more extended than
their light distribution. The cluster discussed here turned out to be
a relatively difficult case, since it was not particularly rich in
galaxies, its relatively low velocity dispersion does not give very
strong image distortions, and the two other projected structures in
the field complicate the analysis. The cluster has a velocity
dispersion of 775 \kms\ and a virial radius of 0.32\hmpc. The virial
radius is somewhat smaller than a comparable $z=0$ cluster, but within
our field the cluster becomes ``clumpy'' and has a steep (logarithmic
slope $>-3/2$)
density profile beyond this radius. The $M/L$ ratio, K corrected and
allowing a luminosity evolution of 0.3 magnitudes, is $255h$, with a
1-$\sigma$ error of 30\%.  The center of light (and X-ray emission)
is close to the cD galaxy, which is about 60\arcs\ removed from the
center of lensing mass. This significance level of the difference in
the apparent luminosity and mass peaks is unclear.  The average
tangential ellipticity calculated using the cD as the center of
symmetry and taking the lensed objects to be at $z_s=0.6$ finds that
the gravitational mass indicated by weak lensing is, between 2 and 3
virial radii, at least twice the virial mass.

What would it take to bring the virial mass and the lensing mass into
agreement? Doubling the cluster $M/L_V(0)$ to about 500$h$ would
remove the discrepancy, but this M/L is less than 0.1\% probable and
would imply an unusually high virial
$\Omega\simeq0.4$. Normally cluster velocity dispersions are
underestimated in small velocity samples.  A concern in this study is
the difference between the peak of the light distribution and the
inferred mass distribution, which is statistically significant, but
subject to various poorly constrained systematic errors. For instance,
the cluster may not be dynamically well relaxed-- the cD is displaced
from the region of greatest galaxy density to the SE, and the galaxies
to the W having somewhat higher redshifts. However, the cD's velocity
is an insignificant 150 \kms\ greater than the cluster mean velocity.
Clusters that are out of dynamical equilibrium normally have
virial masses higher than their true mass, which would increase the
difference between the ``relaxed'' virial mass and the lensing
mass. Therefore the difference is not likely to be found in the virial
analysis.  The mass inferred from lensing depends on the background
galaxies being distributed with the redshifts deduced from field
surveys and the increase in numbers with depth.  If the weighted
$z_s=3$ then the inferred mass is halved from the $z_s=0.6$ assumed
here, however such a high redshift seems quite unlikely in samples
limited at $I=22$ and 23. Calculating distances with a low $\Omega$ of
in a very clumpy universe gives about a 10\% cluster mass
reduction. The analysis certainly could be refined if the source
galaxies had individually known redshifts.  The redshift diagram,
Figure~1, illustrates that the redshift distribution is extremely
clumpy in a field this small, suggesting that the background galaxies
may be similarly distributed.  Nevertheless, the observed distortions,
about 6\% at nearly 0.7\hmpc, where the expected values are 2\% are
far too large to be a result of the virial mass alone.

\ls
\ni\ub{ACKNOWLEDGEMENTS} We thank Greg Fahlman, Nick Kaiser, Gordon
Squires and David Woods for many interesting conversations. The
observations reported here are a small subset of the CNOC cluster mapping
project, whom we thank for assistance and use of these data. Chris Pritchet,
Roberto Abraham and Tammy Smecker-Hane were an invaluable part of the
observing team.

\ifapj \vfill\eject \else\bigskip\fi
\goodbreak
\ls\ls
\ni\ub{REFERENCES}
\parskip=0pt

\ls
\hi{Bardeen, J. M., Steinhardt, P. J. \& Turner, M. S. 1983, Phys. Rev. D., 28,
679}
\hi{Bean, A. J., Efstathiou, G., Ellis, R. S., Peterson, B. A. \& Shanks, T.
1983, \mn 205, 605}
\hi{Bertschinger, E., Dekel, A., Faber, S. M., Dressler, A. \& Burstein, D.
1990, \apj  364, 370}
\hi{Carlberg, R. G. \& Dubinski, J. 1991, \apj 369, 13}
\hi{Carlberg, R. G. 1994, \apj submitted}
\hi{Carlberg, R. G. et al. 1994, JRASC February, in press}
\hi{Coleman, G. D., Wu, C. C., and Weedman, D. W. 1980, \apjs 285, 426}
\hi{Davis, M. \& Peebles, P. J. E. 1983, \apj {267}, 465}
\hi{Fahlman, G., Kaiser, N., Squires, G. \& Woods, D. 1994, preprint}
\hi{Fisher, K. B., Davis, M., Strauss, M. A., Yahil, A., \& Huchra, J. P.
	1993, preprint}
\hi{Gioia, I. M., Maccacaro, T., Schild, R. E., Wolter,A., Stocke, J. T.,
	Morris, S. L., \& Henry, J. P. 1990, \apjs 72, 567}
\hi{Gioia, I. M. \& Luppino, G. A. 1994, \apjs submittted}
\hi{Guth, A. 1981, Phys. Rev. D., 23, 347}
\hi{Kaiser, N., Efstathiou, G., Ellis, R., Frenk, C., Lawrence, A.,
Rowan-Robinson, M., \& Saunders, W. 1991, \mn 252, 1}
\hi{Kaiser, N. \& Squires, G. 1993, \apj 404, 441}
\hi{Kent, S. \& Gunn, J. E., 1982, \aj 87, 945}
\hi{Lynden-Bell, D., Lahav, O., \& Burstein, D. 1989, \mn 241, 235}
\hi{Miralda-Escude, J. 1991, \apj 370, 1}
\hi{Nusser, A. \& Dekel, A. 1993, \apj 405, 437}
\hi{Peebles, P.~J.~E. 1976, \apjl 205, L109}
\hi{Seebok, W.~L. 1986, \apjs 62, 301}
\hi{Schneider, P., Ehlers, J., \& Falco, E. E. 1992 {\it Gravitational Lenses},
(Springer-Verlag: New York)
\hi{Strauss, M. A., Yahil, A., Davis, M., Huchra, J. P., \& Fisher, K. 1992,
	\apj 397, 395}
\hi{Tyson, J. A., Valdes, F., \& Wenk, R. A. 1990, \apjl 349, L1}
\hi{West, M. J., \& Richstone, D. O. 1988, \apj 335, 532}
\hi{Yee, H.~K.~C., Ellingson, E., Carlberg, R.~G, \& Pritchet, C.~J.
1994, in preparation}
\hi{Yee, H.~K.~C. 1991, PASP 103, 396}

\ifapj

\vfill \eject
\centerline{FIGURE CAPTIONS}

\bigskip
\copy11

\bigskip
\copy12

\bigskip
\copy13
\bigskip

\vfill \eject
\centerline{\caps Fig. 1}
\vskip 2.0truecm
\epsfysize=7.5truein
\centerline{\epsffile{zE1224.PS}}
\vfill\eject
\centerline{\caps Fig. 2}
\vskip 2.0truecm
\bigskip
\epsfysize=7.5truein
\centerline{\epsffile{map.PS}}
\vfill\eject
\centerline{\caps Fig. 3}
\vskip 2.0truecm
\epsfysize=7.5truein
\centerline{\epsffile{dis.PS}}
\fi

\bye